\documentclass[twocolumn]{aastex701}

\newcommand{\mytilde}{\raise.19ex\hbox{$\scriptstyle\sim$}}

\newif\ifshowedits
\showeditsfalse 

\ifshowedits
  \newcommand{\edit}[1]{\textcolor{red}{\textbf{#1}}}
\else
  \newcommand{\edit}[1]{#1} 
\fi

\usepackage[utf8]{inputenc}
\usepackage{CJK}
\usepackage{amsmath}
\usepackage{url}
\usepackage{lineno}
\usepackage{float}
\usepackage{placeins}
\usepackage{chngcntr}
\usepackage{bm}
\usepackage{hyperref}

\shorttitle{XLSSC~122: An Active Merger at Cosmic Noon}
\shortauthors{Scofield et al.}

\received{}
\revised{}
\accepted{}

\submitjournal{ApJL}

\graphicspath{{./}{figures/}}

\begin{document}

\title{An Active Galaxy Cluster Merger at Cosmic Noon Revealed by JWST Weak Lensing and Multiwavelength Probes}

\author[orcid=0009-0009-4086-7665,gname=Zachary Pierce, sname=Scofield]{Zachary P. Scofield}
\affiliation{Department of Astronomy, Yonsei University, 50 Yonsei-ro, Seoul 03722, Korea}
\email{zpscofield@yonsei.ac.kr}  

\author[orcid=0000-0002-4462-0709,gname=Kyle, sname=Finner]{Kyle Finner} 
\affiliation{IPAC, California Institute of Technology, 1200 E California Blvd., Pasadena, CA 91125, USA}
\email{kfinner@ipac.caltech.edu}

\author[orcid=0000-0001-9139-5455,gname=Hyungjin, sname=Joo]{Hyungjin Joo} 
\affiliation{Department of Astronomy, Yonsei University, 50 Yonsei-ro, Seoul 03722, Korea}
\email{gudwls4478@gmail.com}

\author[orcid=0000-0002-5751-3697,gname=Myungkook James, sname=Jee]{M. James Jee} 
\affiliation{Department of Astronomy, Yonsei University, 50 Yonsei-ro, Seoul 03722, Korea}
\affiliation{Department of Physics and Astronomy, University of California Davis, One Shields Avenue, Davis, CA 95616, USA}
\email{mkjee@yonsei.ac.kr}

\author[orcid=0000-0002-1566-5094, gname=Wonki, sname=Lee]{Wonki Lee}
\affiliation{Department of Astronomy, Yonsei University, 50 Yonsei-ro, Seoul 03722, Korea}
\email{wonki.lee@yonsei.ac.kr}

\author[orcid=0000-0001-7148-6915,gname=Sangjun, sname=Cha]{Sangjun Cha} 
\affiliation{Department of Astronomy, Yonsei University, 50 Yonsei-ro, Seoul 03722, Korea}
\email{sang6199@yonsei.ac.kr}


\author[orcid=0000-0003-2776-2761,gname=Jinhyub, sname=Kim]{Jinhyub Kim} 
\affiliation{Department of Physics, University of Oxford, Denys Wilkinson Building, Keble Road, Oxford OX1 3RG, UK}
\email{jinhyub.kim@physics.ox.ac.uk}

\author[0000-0001-8792-3091]{Yu-heng Lin}
\affiliation{IPAC, California Institute of Technology, 1200 E California Blvd., Pasadena, CA 91125, USA}
\email[]{ianlin@ipac.caltech.edu}

\author[0000-0001-7583-0621]{Ranga-Ram Chary}
\affiliation{University of California, Los Angeles, CA 90095-1562, USA}
\email[]{rchary@ucla.edu}

\author[0000-0002-9382-9832]{Andreas Faisst}
\affiliation{IPAC, California Institute of Technology, 1200 E California Blvd., Pasadena, CA 91125, USA}
\email[]{afaisst@caltech.edu}

\author[0000-0003-1954-5046]{Bomee Lee}
\affiliation{Korea Astronomy and Space Science Institute, 776 Daedeokdae-ro, Yuseong-gu, Daejeon 34055, Korea}
\email[]{bomee@kasi.re.kr}

\correspondingauthor{M. James Jee}
\email{zpscofield@yonsei.ac.kr, mkjee@yonsei.ac.kr}

\begin{abstract}

The galaxy cluster XLSSC~122 is a rare system at $z = 1.98$, hosting surprisingly evolved member galaxies when the Universe was only one-third of its present age. Leveraging deep JWST/NIRCam imaging, we perform a weak-lensing analysis and reconstruct the cluster's mass distribution, finding a mass peak that coincides with both the X-ray peak and the position of the brightest cluster galaxy. \edit{We obtain a mass estimate of $M_{200\rm c}=1.6 \pm 0.3\times 10^{14}~M_{\odot}$ and a concentration of $c_{200 \rm c}=6.3 \pm 0.3$ implied by the preferred concentration--mass relation}, in agreement with recent strong-lensing estimates. The high concentration in particular motivates tests against empirical and simulation-derived \edit{concentration--mass} relations. Placing our weak-lensing mass map in the context of Chandra X-ray data, MeerKAT radio imaging, ALMA+ACA/ACT Sunyaev-Zel'dovich (SZ) mapping, and \edit{new JWST intracluster light measurements}, we identify consistent NE--SW elongation across datasets and a pronounced offset along the same axis between the SZ and mass/X-ray peaks, pointing to significant merger activity. XLSSC~122 thus serves as a JWST pilot study for high-$z$ lensing, demonstrating the telescope's unique ability to map cluster mass distributions at $z\sim 2$ and motivating a uniform sample of analogous systems with joint lensing, X-ray, SZ, and radio data to probe cluster assembly at \edit{cosmic noon}.

\end{abstract}

\section{Introduction} 

Galaxy clusters are the most massive gravitationally bound objects in the universe, representing the later stages of evolution in the hierarchical model of structure formation. The subject of this study, XLSSU J021744.1-034536 (hereafter XLSSC~122), is a galaxy cluster discovered in the 25 $\rm deg^2$ XXL X-ray survey \citep{pierre2004JCAP...09..011P} at $z=1.98$ \citep{willis2020Natur.577...39W}, corresponding to a lookback time of 10.5 billion years. At this early epoch, its dynamical state and halo concentration offer powerful diagnostics for our understanding of the cluster’s assembly history\edit{.} Crucially, for massive halos at $z \sim 2$, different \edit{halo concentration--mass ($c$--$M$) relations describing $c(M,z)$} diverge significantly. This divergence has been explored in the literature and may be attributed to systematic or numerical effects \citep[e.g.,][]{child2018ApJ...859...55C, 2019diemer}. \edit{Once such effects are controlled for}, high-redshift clusters such as XLSSC~122 \edit{can} provide stringent, empirical tests of early cluster assembly and halo structure, either corroborating $\Lambda$CDM predictions or highlighting tensions if robust discrepancies arise.


A critical step in assessing both a cluster's dynamical state and halo concentration is reconstructing its mass, of which approximately 85\% is dark matter. Gravitational lensing, \edit{or} the deflection of light by a gravitational potential, provides a powerful observational probe for detecting and characterizing dark matter. When a ray of light passes through a massive gravitational potential, its trajectory is altered to an extent set by the gradient of the potential. The resulting distortions of background galaxy images can then be used to reconstruct the projected mass of a lensing cluster. In the central regions of a cluster, where the surface mass density is high, significant light deflection occurs and strong lensing (SL) distorts background galaxy images into multiple images and extended arcs. At lower mass surface densities, weak lensing (WL) manifests as small, coherent distortions across large ensembles of background galaxies. Since gravitational lensing is insensitive to a cluster's hydrodynamical state, it is especially valuable for studying dynamically unrelaxed clusters, which many at high redshifts tend to be. However, because the WL signal-to-noise (S/N) depends on the effective background galaxy density and their lensing efficiency, measuring WL in high-$z$ clusters has been challenging with marginal detections and difficulty constraining morphology \citep{lombardi2005ApJ...623...42L, jee2009ApJ...704..672J, mo2016ApJ...818L..25M, 2017jeeApJ...847..117J, schrabback2018MNRAS.474.2635S, 2020finnerApJ...893...10F}.

JWST's unprecedented depth and resolution now mitigate these limitations, as exemplified by XLSSC~122. Building on the work of \citet{kim2025ApJ...991..109K}, who performed a WL analysis of this cluster with HST, our deep JWST/NIRCam imaging increases the background-galaxy density by a factor of $\sim\!4$ under similar selections. JWST imaging also reveals previously undetected SL arcs \citep{2025finnerApJ...994L..35F}, tightening constraints on the cluster's inner mass profile and concentration. To test these SL inferences and avoid extrapolation beyond the SL regime, we perform a complementary WL mass reconstruction that independently constrains the concentration and large-scale mass profile, leveraging the denser background-galaxy population.

Gravitational lensing is a powerful tool for probing a cluster’s total mass distribution, but multiwavelength data are essential for characterizing its dynamical state. \citet{marrewijk2024A&A...689A..41V} provide a recent \edit{thermal} Sunyaev-Zel’dovich (SZ) and X-ray--based overview of XLSSC~122, including an assessment of its dynamical configuration and possible merger state. Here\edit{,} we extend this picture by incorporating WL, radio, intracluster light (ICL), and cluster member analyses to complement the existing datasets. Together, these observations map the dark matter, hot gas, and stellar components, yielding \edit{a comprehensive, multiwavelength view} of the assembly and evolution of this rare, high-$z$ cluster.

In Section \ref{sec:2}, we describe the observations, data reduction process, and lensing analysis techniques used in this work. In Section 3, we present the results of this \edit{study}, with Sections 4 and 5 reserved for the discussion and conclusions, respectively. We assume a flat $\Lambda$CDM cosmology with $H_0=70~\rm km~s^{-1}~Mpc^{-1}$, $\Omega_{m} = 0.3$, and $\Omega_{\Lambda} = 0.7$. At the cluster redshift of $z=1.98$, the plate scale is $8.38 ~\rm kpc ~\rm arcsec^{-1}$. Masses are reported as $M_{200\rm c}$, which is the mass within a radius $R_{200\rm c}$ where the average density is 200 times the critical density of the universe at the redshift of the cluster. Unless otherwise stated, all right ascension and declination values in this work are referenced in the ICRS coordinate system.

\section{Observations and Data Reduction}\label{sec:2}

\subsection{JWST Imaging}

XLSSC~122 was observed using JWST/NIRCam on 14 August 2024 as part of Program GO 3950 (PI: K. Finner) using four filters: F090W, F200W, F277W, and F356W. The effective exposure times for the mosaics are approximately 4209~s for F200W and F356W, and 7559~s for F090W and F277W. The cluster lies within module A of the NIRCam detector, with a footprint of $\sim\!5.73~\rm arcmin^2$ in the F200W filter. This filter was selected for our WL analysis based on the study of IR systematics in \cite{finnerb-2023ApJ...958...33F} and on the first WL analysis with JWST \citep{finnera-2023ApJ...953..102F}: first, it offers the best ratio of point-spread function (PSF) FWHM ($0\farcs066$) to native pixel scale ($0\farcs031$); second, as the reddest of the short-wavelength filters, it is the most sensitive to high-$z$ background galaxies \citep{2018lee}. Filters redder than F200W (e.g., F277W or F356W) are observed with the long-wavelength channel of NIRCam, which has a lower spatial resolution and therefore reduces the effective source density for WL shape measurements.

The imaging data were reduced using the \texttt{young-jwstpipe} pipeline \citep{youngjwstpipe}, which augments the default JWST Data Calibration Pipeline \citep{2025jwst_pipeline} with additional calibration steps important for robust lensing analyses. For a more detailed description of the data reduction pipeline, see \cite{scofield2025ApJ...993..226S}. The JWST mosaic images are archived and publicly available on Zenodo (\href{https://doi.org/10.5281/zenodo.16810356}{10.5281/zenodo.16810356}). Additional visualizations and documentation are provided at the project page\footnote{\url{https://kylefinner.github.io/xlssc122}}.

\subsection{Cluster-Member Selection}\label{sec:members}

\begin{figure*}
    \centering
    \includegraphics[width=\linewidth]{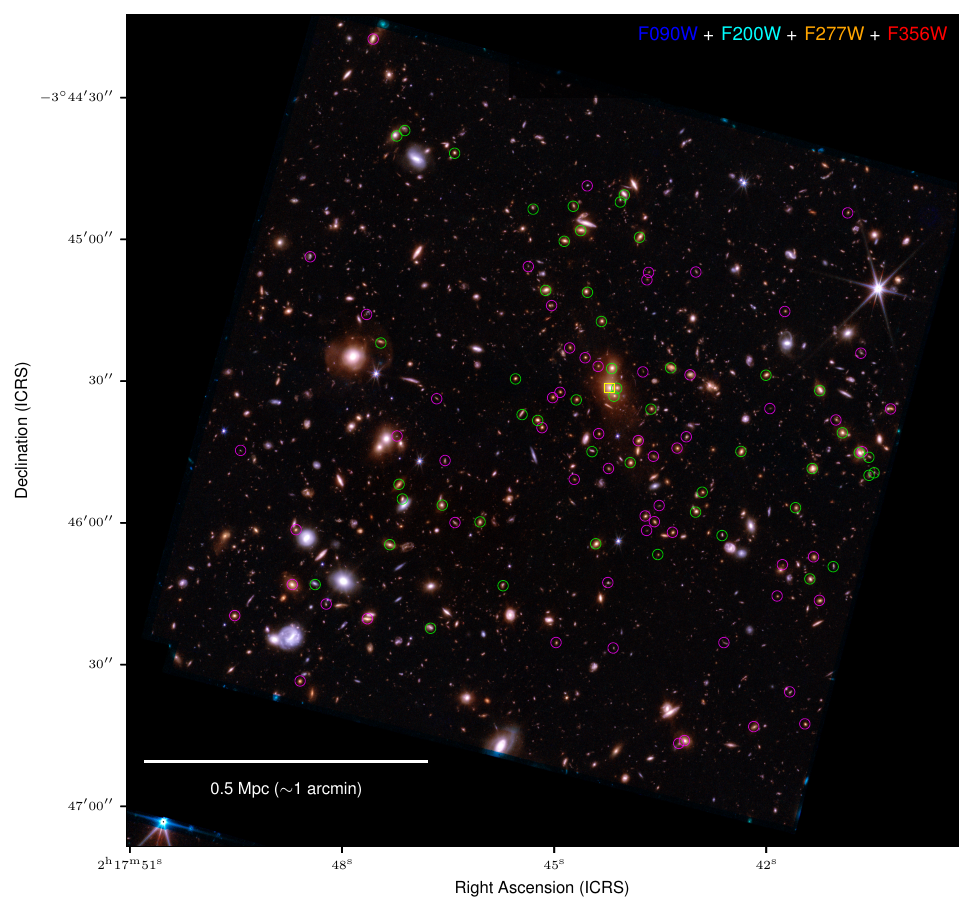}
\caption{Color composite image of JWST/NIRCam module A for the XLSSC~122 field. Spectroscopically selected cluster members (green) and photometrically selected members (magenta) are indicated, and the BCG is marked with a yellow square.}
    \label{fig:cluster_members}
\end{figure*}

We adopt the cluster-member catalog of H. Joo et al., in prep, which combines both spectroscopic and photometric selections and is shown in Figure~\ref{fig:cluster_members}. Spectroscopic member galaxies identified in \cite{willis2020Natur.577...39W}, along with additional members determined using HST grism data, are circled in green. The brightest cluster galaxy (BCG) is marked with a yellow square. Additional cluster member candidates are selected based on photometric redshift (photo-$z$) estimates and highlighted in magenta. This photometric selection was \edit{performed} by combining the four JWST/NIRCam bands with HST imaging in F814W, F105W, and F140W to derive photo-$z$s using \texttt{EAzY-py} \citep{brammer2008ApJ...686.1503B}. We required the entire $1\sigma$ photo-$z$ interval to lie within $\pm25\%$ of the cluster redshift $z=1.98$. This relatively broad range is necessitated by the limited photometric coverage, which inflates photo-$z$ uncertainties. Broader multi-band coverage would reduce these uncertainties and improve the reliability of our member and background galaxy selection, \edit{as} discussed in Section~\ref{sec:foreground}.

\subsection{Multiwavelength Observations}

Multiwavelength observations probe different physical components of the cluster environment. X-ray emission traces the thermodynamic structure of the hot \edit{intracluster medium}, SZ observations measure its integrated thermal pressure and can indicate merging activity, and radio observations reveal diffuse synchrotron emission from relativistic particles in intracluster magnetic fields that often trace turbulence and shocks from merger activity or active galactic nuclei (AGN). Together with accurate mass \edit{reconstructions} from SL and WL, these observations enable a detailed investigation of the dynamical state of XLSSC~122.

The X-ray data used in this work were obtained with Chandra ($0.5$--$7~\rm keV$) and previously presented by \edit{\citet{mantz2018A&A...620A...2M}}. The radio emission map is taken from the MeerKAT International GHz Tiered Extragalactic Exploration survey \citep[MIGHTEE;][]{2025hale, jarvis2016mks..confE...6J}, which performed 1.28 GHz L-band observations on the XMM-LSS field. We use the ALMA+ACA SZ Compton-$y$ map produced by a joint ALMA/ACA–ACT DR6 analysis \citet{marrewijk2024A&A...689A..41V}, which fits the signal simultaneously to ALMA+ACA Band-3 visibilities ($\sim\!$90--105 GHz) and ACT DR6 (100/150) GHz maps \citep{coulton2024PhRvD.109f3530C}. This yields a scale-coupled SZ reconstruction that leverages the complementary spatial-frequency coverage of the interferometric (ALMA/ACA) and wide-field (ACT) data. 

\section{Weak-Lensing Analysis}\label{sec:3}

In WL analyses, dense samples of background galaxy shapes are used to reconstruct the two-dimensional convergence, $\kappa$ (the dimensionless surface mass density), of the lens; here, the galaxy cluster XLSSC~122. The convergence $\kappa$ produces isotropic focusing of background galaxy images, whereas the lensing shear $\bm \gamma$ quantifies anisotropic distortion. Coherent shape distortions induced by the foreground potential are described by the \edit{reduced} shear:
\begin{equation}
    {\bm g} = g_1 + ig_2 = \frac{\bm \gamma}{(1-\kappa)} \, .
\end{equation}
In the WL regime ($\kappa \ll 1$), the measured galaxy ellipticity $\bm e$ provides an approximately unbiased estimator of $\bm g$ after averaging over intrinsic shape noise. The convergence map (or mass map) is then obtained by inverting the shear field with appropriate filtering/regularization (see Section~\ref{sec:massrecon}).

For brevity, we refer the reader to \cite{scofield2025ApJ...993..226S} for a concise overview of the lensing \edit{formalism applied} in our work, and to \cite{Bartelmann:1999yn, Schneider2006} for comprehensive descriptions of gravitational lensing.

\subsection{Background Source Selection}\label{sec:sources}

\begin{figure}
    \centering
    \includegraphics[width=\linewidth]{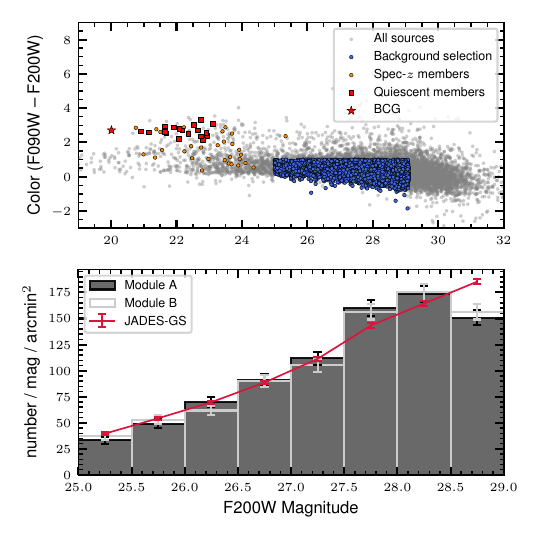}
    \caption{\edit{Color--magnitude diagram for the JWST XLSSC~122 field and comparison of source number densities with a control field.} \textbf{Top:} \edit{Color--magnitude diagram, with all sources identified in a four-filter combined detection image (F090W+F200W+F277W+F356W) shown in gray.} Spectroscopically confirmed cluster members are indicated in orange, \edit{while} quiescent members highlighted with red squares. The BCG and all quiescent members are spectroscopically confirmed members of the cluster. Galaxies selected as background sources for the lensing analysis are marked in blue, and the BCG is \edit{highlighted} with a red star. \textbf{Bottom:} Comparison of number density distributions binned by $m_\mathrm{F200W,\,AUTO}$. The JADES-GS control field is consistent with both modules in the XLSSC~122 field, with the significant deviation in the final magnitude bin attributed to significantly deeper limiting magnitudes ($m_\mathrm{F200W,\,AUTO} \gtrsim 30$) in portions of the control field.}
    \label{fig:cmd}
\end{figure}

Ideally, the WL source catalog should only include galaxies behind the cluster, as foreground or cluster member galaxies are not lensed by the cluster potential and thus dilute the WL signal. However, if we base our source selection on photo-$z$ estimations, there is significant foreground contamination risk from the large photo-$z$ uncertainty discussed in Section~\ref{sec:members}. Such selections are also susceptible to catastrophic redshift outliers that, if present, can bias WL mass estimates \citep{schrabback2021MNRAS.505.3923S}. Therefore, we adopt a simple color $(m_\mathrm{F090W,\,ISO}-m_\mathrm{F200W,\,ISO})$ versus magnitude ($m_\mathrm{F200W,\,AUTO}$) relation to select background sources. The color--magnitude diagram (CMD) is provided in the top panel of Figure~\ref{fig:cmd}. All detected sources are shown in gray, spectroscopically confirmed members of the cluster are marked in orange, quiescent members identified by \citet{2021noordeh} are shown with red squares, and galaxies selected as background sources are plotted in blue. To suppress contamination from cluster members and foregrounds, we select background candidates with $-5 < \rm color < 1$ and $25 < m_\mathrm{F200W,\,AUTO} < 29.1$. The bright-end magnitude limit corresponds to the faintest of the spec-$z$ members, aside from a single outlier at $m_\mathrm{F200W,\,AUTO} \approx 25.3$. The faint magnitude limit is set by the F200W $5\sigma$ depth \citep{2025finnerApJ...994L..35F}. The lower color bound removes a small number of artifacts with unphysical colors, while the upper color bound ensures that all selected galaxies remain bluer than the quiescent member population, thereby reducing contamination from cluster members.

To assess residual contamination, we compare \edit{the F200W magnitude-binned number density} (per $m_{\rm F200W}$ per $\rm arcmin^2$) in the XLSSC~122 field with \edit{that in the JWST Advanced Deep Extragalactic Survey GOODS--South (JADES--GS) field} \citep{merlin2024A&A...691A.240M} in the bottom panel of Figure~\ref{fig:cmd}. \edit{Given that the JADES--GS field lacks a massive cluster, it is used as a control field to establish a baseline expectation for background galaxy counts.} JADES–GS was chosen \edit{as} it provides F090W and F200W imaging with a $5\sigma$ limiting magnitude comparable to our data \edit{($m_{5\sigma}\gtrsim29.1$)}. To ensure a uniform comparison, we mask JADES--GS subregions that do not reach this depth and measure the counts using the same selection and binning as for \edit{the} XLSSC~122 \edit{field}. We find that our source magnitude density shows no excess with respect to the control field at the bright end, suggesting \edit{that cluster member contamination is minimal}. \edit{This control field is also used to derive an effective source redshift ($z_{\rm eff} = 2.384$) based on our source selection criteria, along with the corresponding lensing efficiency ($\langle \beta \rangle = 0.095$) and its width ($\langle \beta^2 \rangle = 0.024$). The latter accounts for the finite width of the source redshift distribution when computing lensing quantities under the single source-plane approximation \citep{2017jeeApJ...847..117J}.}

Finally, to ensure that we are not losing potential lensing constraints, we performed an additional WL analysis using only sources fainter than our chosen magnitude limit ($m_\mathrm{F200W,\,AUTO} > 29.1$). We found that 2064 of these 2789 sources ($\sim\!74\%$) fail the shape measurement stage, and those remaining do not produce any measurable cluster lensing signal.

Our final background source catalog, after imposing shape criteria and removing spurious sources manually (see Section~\ref{sec:measure}), contains a total of 4944 galaxies with 2463 in the cluster-centered module A, corresponding to a total source density of $\sim\!431~\rm arcmin^{-2}$ and module A-only density of $\sim\!433~\rm arcmin^{-2}$. This source density is $\sim\!4\times$ higher than seen in the HST imaging of XLSSC~122 using similar selection criteria \citep{kim2025ApJ...991..109K}, highlighting the power of the JWST for high-$z$ cluster lensing studies. We note that our background source selection strategy may result in foreground contamination, which is discussed further in Section~\ref{sec:massdisc}.

\subsection{PSF Modeling}

Given that the magnitude of WL distortions is extremely small, it is important to account for both telescope-induced distortions and bias in galaxy shape measurements. The first of these is captured by the PSF of the telescope, which must be modeled and accounted for to ensure an accurate WL measurement.

In this work, we model the PSF using an empirical principal component analysis (PCA). PCA derives a set of orthogonal basis functions from the observed stars in the field \citep{jee2007PASP..119.1403J}, which are then used to reconstruct the PSF at any position. In JWST imaging, it has been shown that given an adequate number and distribution of stars in the field, the PSF can be accurately modeled directly from the final mosaic image \citep{finnera-2023ApJ...953..102F, Cha_2024, cha2025ApJ...987L..15C, scofield2025ApJ...993..226S}. The \edit{XLSSC~122 F200W} mosaic has 42 high-quality, well-distributed stars\edit{, which is }far fewer than in typical ground-based, wide-field data. However, because the PSF shows little spatial variation across the field, a PCA model with a low-order polynomial interpolation is sufficient. \cite{Cha_2024} and \cite{scofield2025ApJ...993..226S} compare the PCA approach with the simulation-based \texttt{STPSF} model \citep{webbpsf}, which generates PSFs from optical path difference maps measured onboard JWST. We revisit the comparison between these techniques in Appendix~\ref{psfcompare} to validate these approaches for a new observation and contribute to the growing body of analyses benchmarking PSF modeling strategies for the JWST.

We gauge the performance of \edit{the} PSF model by computing the residual complex ellipticity component ($e_1$ and $e_2$) and size measurements between stars reconstructed with the model and the corresponding observed stars. The complex ellipticity components:
\begin{equation} \label{ellipticity}
    e_1 = \frac{a-b}{a+b} \cos(2\phi) \;, \quad e_2 = \frac{a-b}{a+b} \sin(2\phi) \, ,
\end{equation}
along with the size measurement $R$, are measured using quadrupole moments \citep[see][]{jee2007PASP..119.1403J, mandelbaum2014ApJS..212....5M}. In equation \eqref{ellipticity}, $a$, $b$, and $\phi$ are the semi-major, semi-minor, and position angle of the ellipse, respectively, and $(a-b)/(a+b)$ is the scalar ellipticity. The empirical PSF model yields mean residual ellipticities $\langle e_1 \rangle = (-1.0 \pm 6.7) \times 10^{-4}$ and $\langle e_2 \rangle = (0.4 \pm 7.4) \times 10^{-4}$, and a mean residual size $\langle R \rangle = (0.06 \pm 2.88) \times 10^{-3}$ (uncertainties are standard errors on the mean). For JWST imaging in F200W, the typical \edit{PSF ellipticity amplitude} is $|e|\sim 10^{-2}$\edit{. Here, the mean residuals} are consistent with zero\edit{,} and any residual PSF anisotropy is \edit{suppressed to} the $\lesssim10^{-4}$ level\edit{---well below the level that would significantly impact our WL measurements.}

\subsection{Shape Measurement}\label{sec:measure}
We measure the shapes of selected background galaxies using a forward-modeling technique in which a two-dimensional Gaussian profile is fit to the observed surface brightness distribution using the \texttt{MPFIT} \citep{mpfit} minimization algorithm. The semi-major and -minor axes, position angle, and amplitude are free parameters initialized with the values from \texttt{SExtractor} \citep{bertin-1996A&AS..117..393B}, while the background value and centroid for each source are fixed to their \texttt{SExtractor} values. Noise is accounted for using an rms error map computed from the weight map \edit{produced} by the data reduction pipeline. We require that valid sources have an ellipticity of $e < 0.85$, ellipticity uncertainty $\delta e < 0.3$, and \texttt{STATUS}\footnote{\texttt{STATUS} is an \texttt{MPFIT} parameter indicating the stability of the fit; values other than unity typically correspond to unstable or failed fits.} $= 1$. Sources with an ellipticity of $e \gtrsim 0.85$ are usually spurious detections, while large ellipticity uncertainties indicate unreliable measurements. When performing shape measurements, we cut out galaxy stamps from the mosaic image based on \texttt{SExtractor} major axis length. We require each source cutout side length (semi-major axis \texttt{A\_IMAGE}$\times 4 + 25$) to be less than 100 pixels, as larger sources are typically in the foreground. After shape fitting, we perform a final visual inspection of selected sources to remove any remaining spurious sources, such as the diffraction spikes of bright stars or segmented portions of large galaxies that are likely in the foreground. 

The elliptical Gaussian fitting technique is subject to bias, as mentioned in \cite{finnera-2023ApJ...953..102F}. We calibrate $e1$ and $e2$ using the \texttt{SFIT} technique, which was the best performing method in the GREAT3 challenge \citep{mandelbaum2014ApJS..212....5M} and is described in \citet{2013jee} \edit{and} HyeongHan et al. (in prep.). We simulate galaxy images matched to the XLSSC~122 observing conditions and source population, measure their ellipticities, and derive component-wise multiplicative biases $m_1$ and $m_2$. We find $m_1=-0.24$ and $m_2=-0.22$, and correct the measured shears via $g^{corr}_{i} = g^{meas}_i/(1+m_i)$, i.e., multiply the measured ellipticity components $e_1$ and $e_2$ by $f_{e_1} = \edit{(1/0.76)} \approx 1.32$ and $f_{e_2} = \edit{(1/0.78)} \approx 1.28$, respectively. 

The larger correction factors compared to \cite{finnera-2023ApJ...953..102F} likely stem from differences in both the background galaxy density and the source-selection strategy\edit{, with the SMACSJ0723.3--7327 analysis utilizing a photo-$z$-selected catalog with a background source density of $\sim\!215~\rm arcmin^{-2}$.} A higher source density increases blending and crowding effects, which amplify multiplicative bias and necessitate a stronger correction. The \edit{CMD} source-selection strategy used in this work also retains low--S/N galaxies with stronger noise bias \citep{refregier2012MNRAS.425.1951R} \edit{that} would typically lack reliable photo-$z$s and therefore be removed by a photo-$z$ selection. Together, these effects naturally lead to larger bias corrections in our calibration.

\begin{figure*}[ht]
    \centering
    \includegraphics[width=0.48\linewidth]{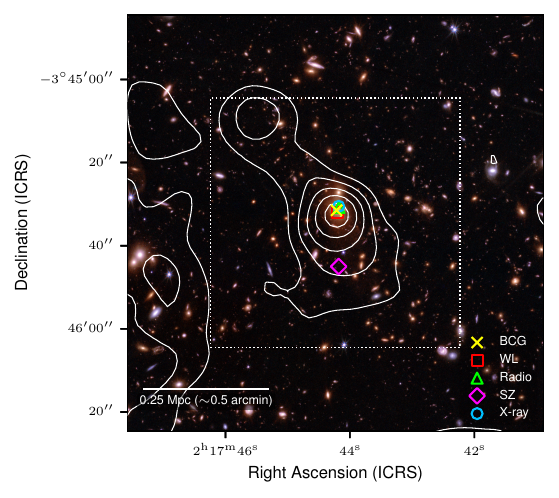}
    \includegraphics[width=0.46 \linewidth]{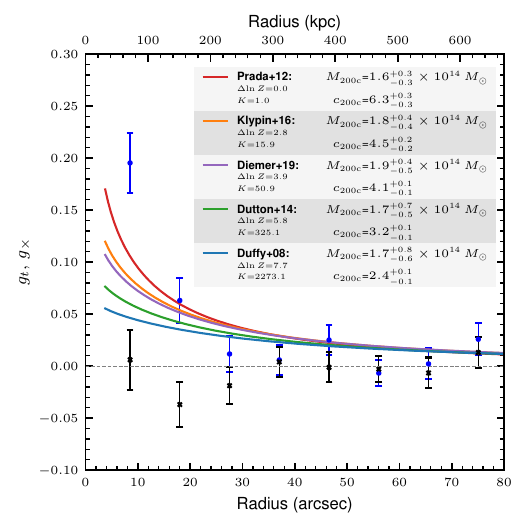}
    \caption{XLSSC~122 WL results. \textbf{Left:} Mass contours produced with \textsc{MARS}. The contours start at the $1\sigma$ level and increase in steps of $1\sigma$. The peak convergence is $\sim\!7.8\sigma$, and the peak convergence contour after smoothing with a Gaussian kernel ($\sigma \approx 3''$) is at $5\sigma$. The BCG is marked with a yellow cross, while the WL, Radio, SZ, and X-ray peak positions are denoted with a red square, green triangle, magenta diamond, and blue circle, respectively. The dotted box marks the region shown in Figure~\ref{fig:matrixplot}. \textbf{Right:} Binned tangential shear (blue dots) and cross shear (black crosses) are shown for visualization purposes. The curves and corresponding labels correspond to \edit{NFW} \edit{profiles fit to the unbinned data} \edit{using} five different \edit{$c$--$M$} relations, with our best-fit model using the \cite{prada2012MNRAS.423.3018P} prescription\edit{.}}
    \label{fig:wl}
\end{figure*}

\section{Results}
\subsection{Mass Reconstruction}\label{sec:massrecon}

We performed the WL mass reconstruction using the free-form MAximum-entropy ReconStruction (\textsc{MARS}) lens-modeling algorithm \citep{Cha_2022, cha_2023ApJ...951..140C, Cha_2024}, which optimizes the convergence on a $100 \times 100$ grid by minimizing a reduced-shear $\chi^2$ objective function regularized with a maximum-entropy (ME) prior: 
\begin{equation}
    f = w\chi^2_{\rm WL} + rR_{\rm ME} \, .
\end{equation}
Here, $\chi^2_{\rm WL}$ represents the $\chi^2$ term for the WL shear, $R_{\rm ME}$ is the ME regularization \edit{favoring the minimally assumptive convergence map consistent with the data}, and the weights $w$ and $r$ determine relative importance. \edit{The weights are fixed a priori and chosen empirically to ensure stable convergence of the minimization.} To assess robustness, we generate 1000 bootstrap resamples of the WL shape catalog and perform the \textsc{MARS} minimization with each. The resulting S/N map is shown in the left panel of Figure~\ref{fig:wl}, with contours starting at the $1\sigma$ level and increasing in $1\sigma$ steps. The mass map is smoothed using a Gaussian kernel with $\sigma \approx 3''$, with a peak contour at $5\sigma$. The unsmoothed S/N map has a peak at $\sim\!7.8\sigma$. 

As a cross-check, we generated a mass map using \mbox{\textsc{FIATMAP}} \citep{wittman2006ApJ...643..128W, wittman2023ApJ...954...36W, stancioli2024ApJ...966...49S}, which implements a classical shear–convergence inversion. The resulting morphology and peak location are consistent with the \textsc{MARS} reconstruction.


\subsection{Mass Estimation}\label{sec:massest}
\begin{figure*}[ht]
    \centering
    \includegraphics[width=\textwidth]{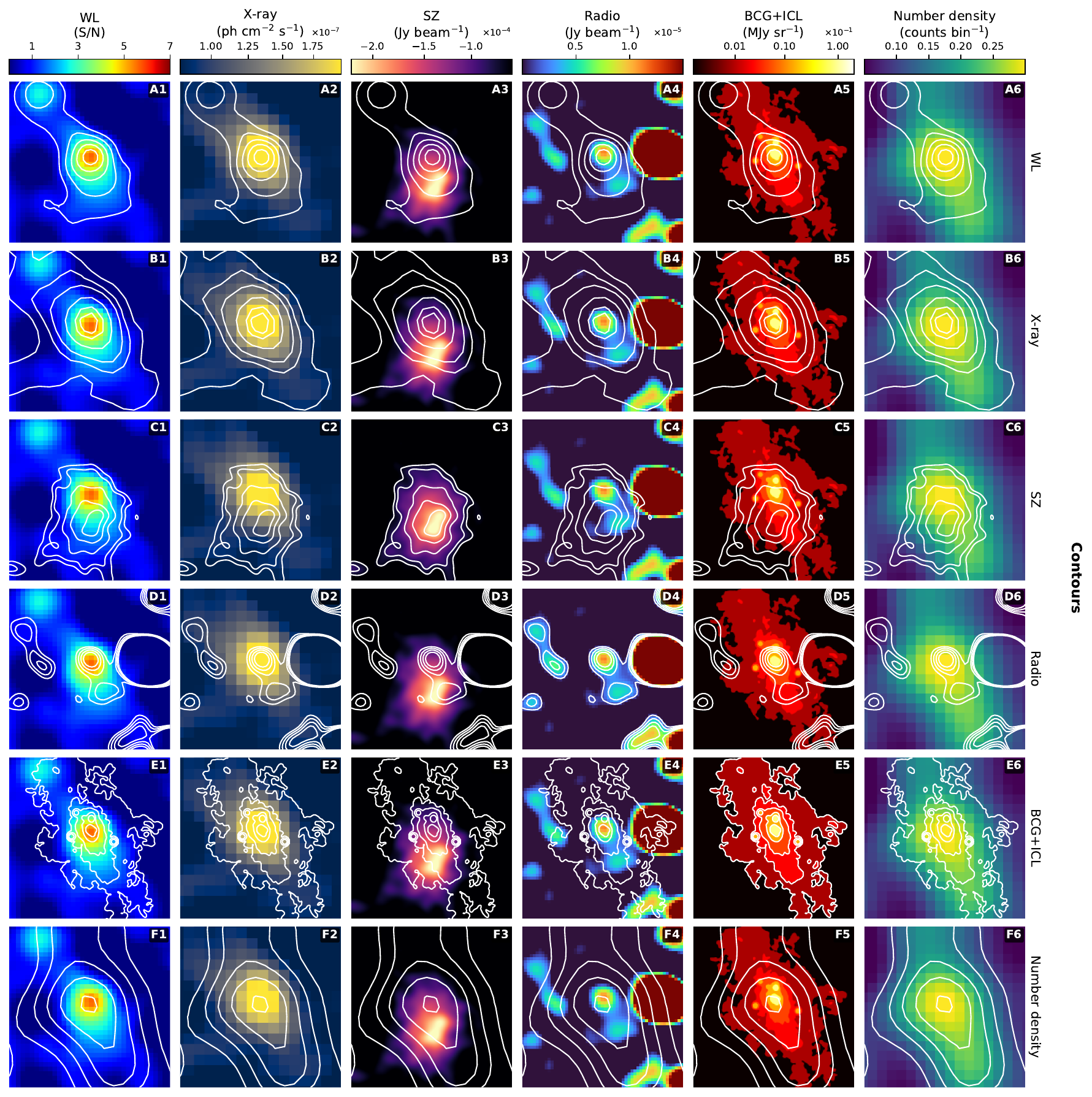}
    \caption{Multiwavelength view of XLSSC~122 within the region marked by the dotted box in Figure~\ref{fig:wl}. Each column displays the following datasets from left to right: WL mass map, Chandra X-ray, MeerKAT radio, ACA+ALMA SZ, BCG+ICL, cluster member number density. Each row shows the corresponding contours for these datasets in the same order from top to bottom\edit{.}}
    \label{fig:matrixplot}
\end{figure*}

Estimating XLSSC~122's mass with a spherical Navarro--Frenk--White \citep[NFW;][]{1997navarroApJ...490..493N} profile fit to the WL shear signal is nontrivial. At $z\simeq2$ the cluster is compact in the JWST field, and the mass contours show a NE--SW elongation consistent with multiwavelength probes (see Figure~\ref{fig:matrixplot}), indicating the system is likely unrelaxed. In such cases, a multi–NFW profile fit could be warranted, but since the inner substructure is unresolved, we adopt a single NFW profile centered on the \edit{BCG}. The right panel of Figure~\ref{fig:wl} shows the binned tangential shear (blue points) and cross shear (black crosses):
\begin{align}
    g_{\rm t} & = -g_1 \cos{2\phi} - g_2 \sin{2\phi} \, , \\
    g_{\times} & = g_1 \sin{2\phi} - g_2 \cos{2\phi} \, ,
\end{align}
where $\phi$ is the polar angle of each source relative to the cluster center. The unbinned tangential shear is used in the NFW \edit{fit}. 

\edit{It is common practice to exclude background sources within an inner radius during \edit{the mass fit} to avoid cluster member contamination, halo centroid uncertainty, and nonlinearity \citep{sommer2025MNRAS.538L..50S,kim2025ApJ...991..109K}. The last of these effects is especially important, as WL measurements are formally valid only in subcritical regions, where the reduced shear satisfies $g \approx \gamma$. Nonlinear lensing distortions near and inside the SL regime cause this approximation to break down, complicating the interpretation of galaxy shapes.}

\edit{To avoid the nonlinear SL regime, we derive the effective Einstein radius ($\theta_E$) for our WL source population using the SL lens model from \cite{2025finnerApJ...994L..35F}. This radius is defined as that of a circle with the same area as that enclosed by the critical curve for a given source redshift ($A_c$):
\begin{equation}
    \theta_E = \frac{1}{d_{\rm L}}\sqrt{\frac{A_c}{\pi}}
\end{equation}
where $d_{\rm L}$ is the angular diameter distance to the lens plane \citep{redlich2012A&A...547A..66R, meneghetti2017MNRAS.472.3177M}. For the critical curve corresponding to our effective source redshift ($z_{\rm eff} = 2.384$), we compute an effective Einstein radius of $\theta_E = 2\farcs4$ and require that all WL sources lie outside this radius. Moreover, to keep the WL inference independent of the SL modeling, we ensure that all identified multiple images are excluded from the WL source catalog.}

\edit{Limited inner profile constraints,} possible foreground contamination, and the use of a single effective lensing efficiency (rather than per-source redshifts) reduce our ability to constrain $M_{200\rm c}$ and $c_{200\rm c}$ with uninformative priors. \edit{When both parameters are fit simultaneously, the posterior favors an implausibly high concentration ($c_{200\rm c} \gtrsim 20$) and a correspondingly lower mass ($M_{200\rm c} \sim 7\times10^{13}~M_{\odot}$), reflecting the well-known $c$--$M$ degeneracy resulting from limited inner-radius constraints and source-population redshift assumptions. This mass, estimated simultaneously with the concentration, is well below the existing WL, SL, and SZ inferences for this system \citep{kim2025ApJ...991..109K, 2025finnerApJ...994L..35F, marrewijk2024A&A...689A..41V}.} 

We therefore fit only $M_{200\rm c}$ and test five \edit{$c$--$M$} prescriptions \citep{prada2012MNRAS.423.3018P, klypin2016MNRAS.457.4340K, 2019diemer, 2014dutton, 2008duffy}, as shown in the right panel of Figure~\ref{fig:wl}. We rank the \edit{$c$--$M$} prescriptions by the Bayesian evidence $Z$, computed with \edit{\texttt{PyMultiNest}} \citep{multinest2009MNRAS.398.1601F, buchner2014A&A...564A.125B, Buchner2016S&C....26..383B} via nested sampling. The evidence is the prior-weighted likelihood,
 \begin{equation}
     Z \equiv p(D|M)=\int \mathcal{L}(\bm \theta) \pi(\bm \theta) d\bm \theta \, ,
 \end{equation}
where $p(D|M)$ is the probability of the data $D$ given the model (\edit{$c$--$M$} prescription) $M$, $\bm{\theta}$ denotes the model parameters, $\mathcal{L}(\bm{\theta})$ is the likelihood, and $\pi(\bm{\theta})$ is the prior distribution. The Bayes factor between models $M_i$ and $M_j$ is defined as
 \begin{equation}
    K_{ij} \equiv \frac{p(D|M_i)}{p(D|M_j)} = \frac{Z_i}{Z_j} \, ,
\end{equation}
following standard Bayesian model selection theory \citep[e.g.,][]{jeffreys1961, Kass01061995, trotta2008ConPh..49...71T}. With equal model priors $p(M_i) = p(M_j)$, the Bayes factor equals the posterior odds:
\begin{equation}
    K_{ij} = \frac{p(M_i|D)}{p(M_j|D)} \, .
\end{equation}
Thus, the Bayes factor $K_{ij}$ quantifies how strongly the data favor one \edit{$c$--$M$} prescription over another.

It is often useful to express the Bayes factor in logarithmic form:
\edit{
\begin{align}
    \Delta \ln Z_{ij} & \equiv \ln Z_i - \ln Z_j , \\
    \qquad K_{ij} & = \exp\!\big(\Delta \ln Z_{ij}\big).
\end{align}
}
According to the Kass-Raftery scale \citep{Kass01061995}, \edit{$\Delta \ln Z \approx 1\text{--}3 \ (K\sim 3\text{--}20)$} is ``positive,'' \edit{$\Delta \ln Z \approx 3\text{--}5 \ (K\sim 20\text{--}150)$} is ``strong,''  and \edit{$\Delta \ln Z \gtrsim 5 \ (K \gtrsim 150)$} is ``very strong'' evidence for model $M_i$ over $M_j$. The right panel of Figure~\ref{fig:wl} lists \edit{$\Delta \ln Z$} and the corresponding Bayes factor $K$ for the \cite{prada2012MNRAS.423.3018P} prescription ($M_i$) and alternative prescriptions ($M_j$). Alternatives are disfavored by \edit{$\Delta \ln Z \approx 2.8\text{--}7.7$}, with corresponding Bayes factors $K \approx 16\text{--}2273$, indicating positive (near-strong) to very strong support for \cite{prada2012MNRAS.423.3018P} over the other models. Adopting this relation, we infer a mass of $M_{200\rm c}=1.6 \pm 0.3\times 10^{14}~M_{\odot}$ and an implied concentration of $c=6.3 \pm 0.3$. These values agree with the independent SL constraints of \cite{2025finnerApJ...994L..35F} at the $1\sigma$ level and are discussed further in Section~\ref{sec:massdisc}.

\subsection{Multiwavelength Context}\label{sec:multi}

Figure~\ref{fig:matrixplot} places the WL mass reconstruction in the context of multiwavelength probes of XLSSC~122. In order, each column (labeled 1--6) shows the WL S/N, X-ray, SZ, radio, BCG+ICL, and cluster member number density map. Each row, labeled A--F from top to bottom, shows the corresponding contours in the same order as the columns. This format allows for direct comparisons between pairs of datasets, revealing \edit{the relative alignment and spatial offsets among the various tracers of the cluster.}

\medskip
\noindent\textbf{Morphological consistency between datasets:}
We find that the BCG, WL peak, X-ray peak, and radio peaks are all highly consistent, \edit{as illustrated} in the left panel of Figure~\ref{fig:wl}. Their separation is smaller than the radio beam FWHM ($8\farcs9$), the rebinned Chandra map pixel scale ($\sim\!3\farcs94~\rm pixel^{-1}$), and at most comparable to the WL grid resolution ($\sim\!1\farcs7~\rm pixel^{-1}$). \edit{In addition, we} observe a consistent NE--SW elongation and position angle among these datasets \edit{which is also apparent} in the BCG+ICL (column 5) and cluster member number density (column 6) maps \edit{in Figure~\ref{fig:matrixplot}}. 

\medskip
\noindent\textbf{WL--SZ peak offset:}
In contrast, the SZ distribution shows a $\sim\!12$--$14''$ ($\sim\!100$--$117~\rm kpc$) offset from the WL mass peak (see Figure~\ref{fig:wl} and panels A3 and C1 of Figure~\ref{fig:matrixplot}), with this range defined by the coarser WL mass map. \cite{kim2025ApJ...991..109K} report a $\sim\!7''$ offset between their WL mass peak and SZ peak, with this separation being within their $1\sigma$ WL mass centroid uncertainty. Using the SZ synthesized beam (BMAJ $\times$ BMIN = $3\farcs36 \times 2\farcs21$; circularized to $2\farcs7$), our offset is $\sim\!4.4$ times the SZ beam FWHM. Even after applying the $5''$ taper from \cite{marrewijk2024A&A...689A..41V}, the offset remains $\sim\!2.4\times$ larger than the effective beam size. These factors argue against a stochastic or processing-related origin; instead, the offset likely reflects cluster dynamics that displace the thermal-pressure (SZ) peak from the mass (WL) peak. This is discussed further in Section~\ref{sec:merger}. 

\medskip
\noindent\textbf{Diffuse radio emission:}
The radio emission (column 4) at XLSSC~122's position is a 2.5$\sigma$ peak \edit{excess} in the MIGHTEE 8\farcs9 resolution radio data. Since this emission is not detected as a point source at higher resolution (5\farcs5), we propose that it may be diffuse emission, such as a radio halo, originating from the cluster. The detection of a radio halo at $z \approx 2$ would have significant implications for the evolution \edit{timescale} of magnetic fields in galaxy clusters. However, the low S/N, low resolution, and lack of spectral index hinder our ability to identify the nature of this low-significance feature. 

\medskip
\noindent\textbf{ICL--mass alignment:}
In Joo et al. (in prep), we compare the spatial distribution of the BCG+ICL measured in the F356W image (Figure~\ref{fig:matrixplot}, column 5) with the SL mass map \citep{2025finnerApJ...994L..35F}.
The two distributions show a high degree of similarity in the central region but diverge at projected radii of $\sim\!100 ~\rm kpc$.
We quantify the spatial similarity between the ICL and SL mass distributions using \texttt{pyWOC} \citep{Yoo2022}, where values range from 0 (fully dissimilar) to 1 (identical).
The similarity is highest near the center ($\sim\!0.8$) and declines to $\sim\!0.68$ at a radius of $\sim\!100~\rm kpc$.
When compared with the WL mass distribution derived in this work (panels A5 and E1), the similarity improves to $\sim\!0.75$ at the same radius.
The Modified Hausdorff Distance \citep[MHD;][]{Dubuisson1994}, which measures the typical boundary-to-boundary separation by averaging nearest-neighbor distances between contours, yields a similar trend: the boundary separation between the SL--ICL contours is $\sim\!58~\rm kpc$, whereas the WL--ICL comparison shows a reduced offset of $\sim\!18~\rm kpc$. This improvement is likely due to WL more accurately probing the outer regions of the cluster, resulting in the reconstruction of an elongated mass structure coinciding with the ICL extension. 

\section{Discussion}

\subsection{\edit{Source Selection and Foreground Contamination}}\label{sec:foreground}

\edit{Tighter WL constraints on the cluster mass and concentration will ultimately require improved selection of background sources.} Our CMD selection effectively limits member contamination, but for a $z\simeq2$ cluster, the risk of foreground contamination is substantial. Using photo-$z$ statistics from the JADES--GS field as a reference, we estimate that our \edit{CMD} cut could admit $\sim\!53\%$ foreground sources. Such contamination dilutes the WL signal and biases the \edit{effective} lensing efficiency low \edit{as a result of down-weighting according to the inferred foreground fraction.} Nevertheless, the $\sim\!7.8\sigma$ WL S/N peak may suggest that contamination is less severe in the XLSSC~122 field than this control field estimate implies, likely in part due to our strict shape-fitting criteria. Robust photo-$z$s for WL sources would both suppress foreground contamination and enable per-source lensing weights, potentially strengthening constraints on the mass and concentration. Additionally, these redshifts would aid in identifying multiple-image systems for SL analyses.

The four available NIRCam filters for XLSSC~122 are insufficient for reliable photo-$z$s at the depths reached, particularly for high-$z$ objects. For future high-$z$ ($z\sim 2$) cluster lensing analyses, we advocate broad JWST coverage from $\sim\!0.9$--$4.4\mu m$: a blue anchor (F090W or HST F606W/F814W) to suppress dusty low-$z$ interlopers and stabilize photo-$z$s; dense sampling at 1.0--1.2~$\mu m$ (F115W + F150W) to bracket the 4000$\rm\AA$ and Balmer breaks for member identification; and red-side leverage to map ICL and constrain SED curvature. 

\subsection{Mass and Concentration Constraints}\label{sec:massdisc}

The preference for the \cite{prada2012MNRAS.423.3018P} \edit{$c$--$M$} prescription in our WL mass fit is consistent with its prediction of higher concentrations for massive halos at $z\sim2$; however, this preference does not require that the prescription yields the \edit{physically} correct mass and concentration for XLSSC~122. In principle, $M_{200\rm c}$ and $c_{200\rm c}$ would be fit jointly with weakly informative priors, but, as discussed in Section~\ref{sec:massest}, the CMD-based source selection and limited radial leverage keep $c_{200\rm c}$ poorly constrained.

\edit{The inner WL exclusion radius used in Section~\ref{sec:massest} ($2\farcs4$) is smaller than the distance between the BCG and the closest neighboring WL source, and therefore there are no sources removed by the $\theta_E$ cut. $\theta_E$ increases as a function of source redshift \citep{meneghetti2017MNRAS.472.3177M}, and it is possible that the effective source redshift estimated from the JADES--GS control field is biased low. This possibility is consistent with the fact that the $z_{\rm eff}$ inferred using the Hubble Ultra Deep Field \citep[HUDF;][]{rafelski2015AJ....150...31R} in \cite{kim2025ApJ...991..109K} matches the value ($z_{\rm eff} = 2.384$) derived from the JADES--GS field, despite the latter field being $\sim\!0.8~\rm mag$ deeper and the present analysis adopting a fainter source selection limit. One potential cause of such a bias is photo-$z$ uncertainty in the JADES--GS control field, which was identified as an issue for low S/N sources in the field \citep{merlin2024A&A...691A.240M}. A second contribution could be the $53\%$ inferred foreground contamination discussed in Section~\ref{sec:foreground}, which down-weights the effective source redshift. Both of these issues can lead to an underestimation of $\theta_E$, and consequently, the inclusion of sources from the nonlinear regime.} 

\edit{In case $z_{\rm eff}$ is biased low, we repeated our analysis assuming $z_{\rm eff} = 3.8$, corresponding to $\theta_E = 8\farcs8$ and matching the critical curve shown in \cite{2025finnerApJ...994L..35F}. This assumption is intentionally conservative, enforcing a more aggressive exclusion of the nonlinear SL regime from the mass estimation. Using $\theta_E = 8.8$ as a WL exclusion radius in the NFW \edit{fit} reduces our ability to differentiate between $c$--$M$ prescriptions statistically, as our constraining power in the inner region is significantly limited. However, even with this limitation, the \cite{prada2012MNRAS.423.3018P} relation is weakly preferred by Bayesian model selection ($K \approx 1.4 \text{--}2.7$) and yields an inferred concentration $c_{200 \rm c} = 5.5 \pm 0.4$.}

\edit{While the data statistically favor higher concentrations, the physical interpretation of this result requires caution. First, the elevated concentrations inferred under the \cite{prada2012MNRAS.423.3018P} relation arise from their use of the maximum circular velocity $V_{\rm max}$ under the assumption of an NFW profile \citep{meneghetti2013arXiv1303.6158M,diemer2015ApJ...799..108D}. In unrelaxed systems, boosts in $V_{\rm max}$ can artificially inflate the inferred concentration when an NFW profile is enforced. Second, the WL mass distribution and multiwavelength data indicate that XLSSC~122 is dynamically active, suggesting that a single spherical NFW halo provides, at best, a simplified description of its mass distribution. Consequently, the preference for the \cite{prada2012MNRAS.423.3018P} relation should not be interpreted physically but instead as an indication that the WL signal favors a relatively steep inner mass profile.}

\subsection{\edit{Large-Scale Structure Covariance}}
\edit{In addition to uncertainties from intrinsic shape scatter and measurement error, uncorrelated large-scale structure (LSS) along the line-of-sight introduces covariance in the observed shapes of background galaxies, which can increase uncertainties in WL mass estimates.}

\edit{To quantify the impact of LSS on our mass estimate, we modeled the LSS contribution using Gaussian random fields drawn from the theoretical convergence power spectrum, $C_{\ell}^{\kappa \kappa}$. This power spectrum was computed using \texttt{pyccl} \citep{pyccl2019ApJS..242....2C} and an empirical source redshift distribution derived from the same JADES--GS control field and photometric selection discussed in Section~\ref{sec:sources}. We generated 1000 realizations of the LSS convergence and reduced shear fields ($g_1$, $g_2$) on periodic flat-sky grids with resolution $8192 \times 8192$ and band-limited in angular scale. The LSS shear field spans angular scales from $57\farcm09$ (corresponding to the full extent of the simulated field) down to $1\farcs2$, consistent with the median nearest-neighbor separation of the source population and above the Nyquist limit of the simulated shear grid. The LSS-induced reduced shear field was then sampled at the observed galaxy positions, added to the measured shapes, and the best-fit cluster mass was re-estimated for each realization using the \cite{prada2012MNRAS.423.3018P} $c$--$M$ relation.}

\edit{The scatter in recovered best-fit mass across the 1000 realizations provides an estimate of the additional uncertainty due to uncorrelated LSS. When added in quadrature with the statistical uncertainty, the LSS term increases the total WL mass uncertainty by $\sim\!31\%$ relative to the statistical uncertainty alone, resulting in a final WL mass estimate of $1.60 \pm 0.30~\text{(stat.)} \pm 0.26~\text{(LSS)} \times 10^{14}~M_{\odot}$. Propagating the LSS-broadened mass uncertainty through the adopted $c$--$M$ relation increases the implied concentration uncertainty by $\sim\!0.1$. For comparison, the increase in WL mass uncertainty here due to LSS is larger than the 15--20\% increase reported in \cite{kim2025ApJ...991..109K}. This likely reflects differences in methodology, details of the LSS power-spectrum construction, and angular-scale treatment.}

\subsection{Merger Evidence and Cosmological Implications}\label{sec:merger}


XLSSC~122 is the highest redshift cluster to be analyzed with WL, and the most distant to exhibit SL features. Our WL analysis favors a high concentration and is consistent with independent SL modeling. \edit{At lower redshifts, apparent cluster overconcentration has been extensively investigated in surveys such as the Local Cluster Substructure
Survey \citep[LoCuSS;][]{okabe2010PASJ...62..811O} and the Cluster Lensing And Supernova survey with Hubble \citep[CLASH;][]{postman2012ApJS..199...25P, umetsu2014ApJ...795..163U, umetsu2016ApJ...821..116U}. These studies demonstrated that projection effects, halo triaxiality, and selection biases can induce systematically elevated lensing-inferred concentrations. When these effects are properly marginalized over, the resulting $c$--$M$ inferences were found to be consistent with $\Lambda$CDM expectations.} 

\edit{In the case of XLSSC~122, the SZ--X-ray peak offset and pronounced plane-of-sky elongation of the WL mass map argue against a purely line-of-sight projection being the dominant cause of elevated concentration. However, given the dynamically disturbed state of the system, triaxiality remains a plausible contributor. Moreover, with a significantly higher redshift than other clusters in the ACT DR5 SZ-selected sample, XLSSC~122 may represent an extreme system preferentially selected from the tail of the population, rather than a representative cluster at cosmic noon.}

\edit{Even if geometric or selection effects contribute to the elevated lensing-inferred concentration, XLSSC~122 may nonetheless be intrinsically concentrated, in which case a high concentration at $z\simeq2$ would point to early halo assembly for this system.} Potential contributors include (i) baryonic effects in the core (e.g., BCG contraction) that steepen the inner total-mass profile and (ii) genuine early collapse \citep{2001bullock, 2002wechsler}. \cite{2025finnerApJ...994L..35F} describe two scenarios: (a) accelerated assembly associated with the cluster's quiescent, `bulge-like' cluster member population \citep{2021noordeh}, and (b) modified early-time expansion histories (e.g., early dark energy) that could shift structure growth earlier. Discriminating between these contributors \edit{at the population level} will require numerous $z\gtrsim1.5$ clusters with WL, SL (if features exist), X-ray, and SZ analyses. As mentioned in Section~\ref{sec:massdisc}, robust photo-$z$s (or ideally, spec-$z$s) will be critical in separating projection and selection effects from population trends. 

The WL and cross-band morphology support ongoing or recent merger activity in XLSSC~122. The mass, X-ray, ICL, radio, and member galaxy density maps shown in Figure~\ref{fig:matrixplot} share a common elongation, consistent with merging activity along a shared axis. Additionally, we find a significant offset between the SZ peak and both the mass and X-ray peaks. SZ--X-ray peak dissociations are naturally produced in cluster mergers as they trace different thermodynamic quantities of the ICM: X-ray surface brightness scales with the gas density squared and is dominated by dense subcluster cores, while the thermal SZ signal scales with the integrated electron pressure ($n_eT_e$) and is enhanced in regions where the ICM is shock-heated or compressed. A merger-driven pressure boost would offer a natural explanation as to why this exceptionally high-$z$ cluster was detected by ACT in DR5, where it stood out relative to the typical redshift distribution of SZ--selected clusters. Indeed, \cite{marrewijk2024A&A...689A..41V} note that the detection could be explained by a subhalo core passage, consistent with a dynamically disturbed state.

Offsets between SZ or X-ray peaks and the BCG are often used to identify merging clusters \citep{zenteno2020MNRAS.495..705Z}. \edit{Moreover, multiwavelength component dissociations and their configuration can be used to further characterize merger states and cluster assembly at cosmic noon \citep[e.g.,][]{andreon2023MNRAS.522.4301A,felix2024MNRAS.534.3676F}.} \edit{We observe a significant offset between the BCG and the SZ peak ($\sim\!100~\rm kpc$) in XLSSC~122}, but no meaningful offset between the BCG and the X-ray peak. To explore the physical origins of this configuration, we investigate multiwavelength component offsets in 200 clusters at $z\approx2$ with masses comparable to XLSSC~122 ($M_{200\rm c}=[0.4,~2]\times10^{14}~M_{\odot}$) in the cosmological zoom-in simulation \textsc{TNG-Cluster} \citep{nelson2024A&A...686A.157N}. We identify two primary mechanisms capable of generating significant BCG--SZ peak offsets without a corresponding BCG--X-ray peak offset: merger activity and AGN feedback. In AGN-driven cases, however, the BCG--SZ peak offsets are substantially smaller (typically $\sim\!50~\rm kpc$) than the separation observed in XLSSC~122. Moreover, \cite{marrewijk2024A&A...689A..41V} reject the presence of an AGN-like point source at the location of the BCG in the ALMA+ACA data at a 4.1$\sigma$ level. We therefore conclude that the SZ peak offset in XLSSC~122 is most likely merger-driven. 

Determining whether XLSSC~122 is in a pre- or post-merger state is more challenging, as both can produce the observed $\sim\!100~\rm kpc$ BCG--SZ peak separation. However, in \textsc{TNG-Cluster} mergers, pre-merger dissociated SZ peaks often coincide with bright member galaxies of infalling groups, whereas the SZ peak in XLSSC~122 is notably isolated from cluster members. Post-merger systems in the \edit{\textsc{TNG-Cluster} sample} more commonly exhibit isolated SZ peaks, a configuration that aligns with what we observe in XLSSC~122. The argument for a post-merger state is further supported by the extended ICL signal in XLSSC~122, which is asymmetric and aligns with the NE--SW axis seen in the WL and X-ray maps. This ICL morphology is naturally produced in merging clusters \citep{Joo2025, cha2025ApJ...987L..15C}. \edit{We show three representative examples from \textsc{TNG-Cluster} illustrating each mechanism capable of driving BCG--SZ offsets without corresponding BCG--X-ray offsets in Appendix~\ref{sec:tng}.}

Finally, in Section~\ref{sec:multi} we report a tentative 2.5$\sigma$ diffuse radio excess at the cluster position. Diffuse radio emission from cluster centers can be classified as AGN activity, \edit{relics (from merger-induced shocks), or radio halos (including mini halos). If \edit{the} diffuse radio excess in XLSSC~122 were confirmed as a radio halo, this would imply that turbulence is present in the ICM and the intracluster magnetic fields are already established at this epoch.} Recent detections of radio halos (and candidates) at $z>1$ have been demonstrated by radio telescopes such as MeerKAT and LOFAR \citep{2025digennaro, 2025sikhosana, 2025hlavacek-larrondo}. However, given the low significance of the radio \edit{feature}, we refrain from classification and note that additional, multi-frequency radio observations are required to confirm or refute a halo interpretation.

\section{Conclusion}

\edit{We present a WL mass reconstruction of the $z=1.98$ cluster XLSSC~122 based on deep JWST/NIRCam imaging.} Bayesian model selection reveals that the \cite{prada2012MNRAS.423.3018P} $c\text{--}M$ relation is preferred over the other prescriptions tested. \edit{Adopting this relation yields a mass of $M_{200 \rm c} = 1.6 \pm 0.3 \times 10^{14}~M_{\odot}$ and an implied concentration of $c_{200\rm c}=6.3 \pm 0.3$, with both of these values} in $1\sigma$ agreement with independent SL constraints. \edit{Tests with and without employing $c$--$M$ prescriptions consistently favor a high concentration, within the simplifying assumption of a single-halo NFW description for this dynamically disturbed system.} In a multiwavelength context, the WL, X-ray, and ICL maps share a common elongation, while the SZ peak is significantly displaced from the mass/X-ray peaks along the shared axis. \edit{Collectively}, these features point to \edit{ongoing merger activity}.

Single-object inferences remain sensitive to projection, baryonic effects in the core, \edit{dynamical state}, and foreground contamination. However, if supported by an ensemble of high-$z$ clusters, high concentrations at this epoch could imply early halo assembly and provide a stringent test of halo-structure prescriptions within $\Lambda$CDM.  Broader JWST photometric coverage enabling per-source redshifts, together with high-resolution SZ and X-ray thermodynamic mapping, will sharpen mass and concentration constraints, clarify merger states, and assess the prevalence of such high-$c$ systems. XLSSC~122 serves as a JWST pilot for high-$z$ lensing and motivates a uniform sample of analogous clusters with joint WL+SL, X-ray, SZ, and radio data, enabling robust measurements of cluster assembly at \edit{cosmic noon}.

\section{Software and third party data repository citations} \label{sec:cite}
\emph{SExtractor} \citep{bertin-1996A&AS..117..393B}; \emph{Colossus} \citep{2018diemer}; \emph{PyMultiNest} \citep{buchner2014A&A...564A.125B, Buchner2016S&C....26..383B}; \emph{pyccl} \citep{pyccl2019ApJS..242....2C}; \emph{eazy-py} \citep{brammer2008ApJ...686.1503B}; emph{JWST Science Calibration Pipeline} \citep{2025jwst_pipeline}; \emph{YOUNG JWST Pipeline} \citep{youngjwstpipe}; \emph{YOUNG STPSF Mosaic} \citep{stpsfmosaic}.

\begin{acknowledgments}
We are grateful to Joshiwa van Marrewijk for providing the SZ data used here. 

This work is based [in part] on observations made with the NASA/ESA/CSA JWST and downloaded from the Mikulski Archive for Space Telescopes (MAST) at the Space Telescope Science Institute (STScI), which is operated by the Association of Universities for Research in Astronomy, Inc., under NASA contract NAS 5-03127 for JWST. These observations are associated with program \#3950. Support for program \#3950 was provided by NASA through a grant from the Space Telescope Science Institute, which is operated by the Association of Universities for Research in Astronomy, Inc., under NASA contract NAS 5-03127.

The JWST data presented in this article were obtained from MAST at STScI. The specific observations analyzed can be accessed \edit{via~\dataset[DOI: 10.17909/hemh-ca02]{https://doi.org/10.17909/hemh-ca02}.}

\edit{This paper employs a list of Chandra datasets, obtained by the Chandra X-ray Observatory, contained in~\dataset[DOI: 10.25574/cdc.525]{https://doi.org/10.25574/cdc.525}.}

MJJ acknowledges support for the current research from the National Research Foundation (NRF) of Korea under the programs 2022R1A2C1003130 and RS-2023-00219959.

Bomee Lee is supported by the National Research Foundation of Korea (NRF) grant funded by the Korean government (MSIT), 2022R1C1C1008695.
\end{acknowledgments}

\begin{contribution}
Z. P. Scofield and K. Finner performed the WL analysis. H. Joo estimated photometric redshifts and carried out the ICL analysis. Z. P. Scofield wrote the manuscript, with revisions from K. Finner, M. J. Jee, and H. Joo. M. J. Jee supervised the graduate students and provided expertise in gravitational lensing. S. Cha developed the lens-modeling algorithm used in this work and contributed to the analysis. W. Lee analyzed the \textsc{TNG-Cluster} simulations, and Y. Lin reduced the HST grism data and derived grism redshifts. J. Kim provided the HST imaging for photometric-redshift estimation. K. Finner secured funding and led the JWST proposal (Program GO~3950). A. Faisst, B. Lee, and R. Chary contributed to both this manuscript and the JWST proposal.
\end{contribution}

\facilities{JWST (NIRCam), HST (WFC3), Chandra, ALMA (ACA), MeerKAT, ACT}

\clearpage

\appendix
\twocolumngrid
\counterwithin{figure}{section}

\section{PSF Model Comparison} \label{psfcompare}

\begin{figure}[!t]
    \raggedleft
    \includegraphics[width=\linewidth]{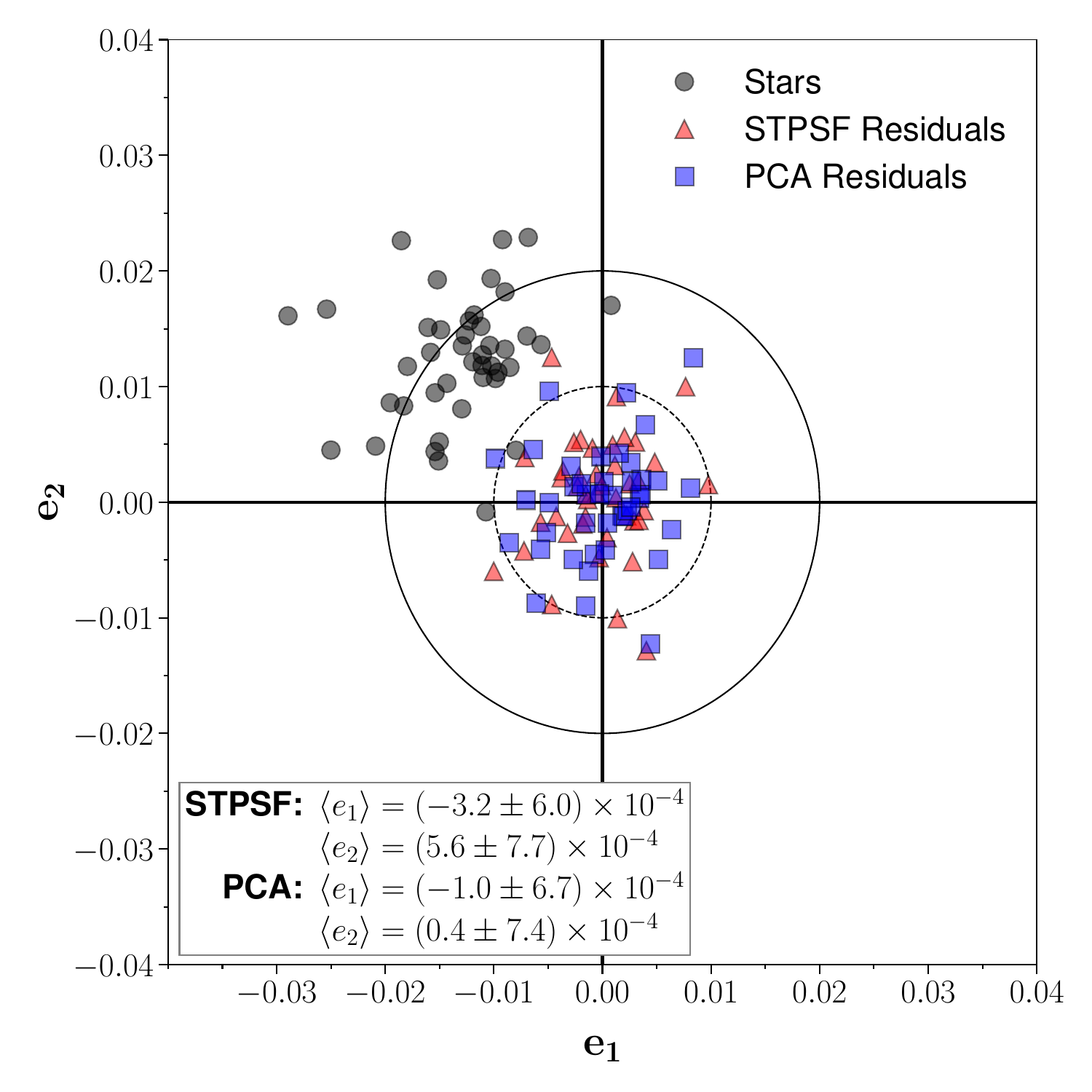}
    \includegraphics[width=0.94\linewidth]{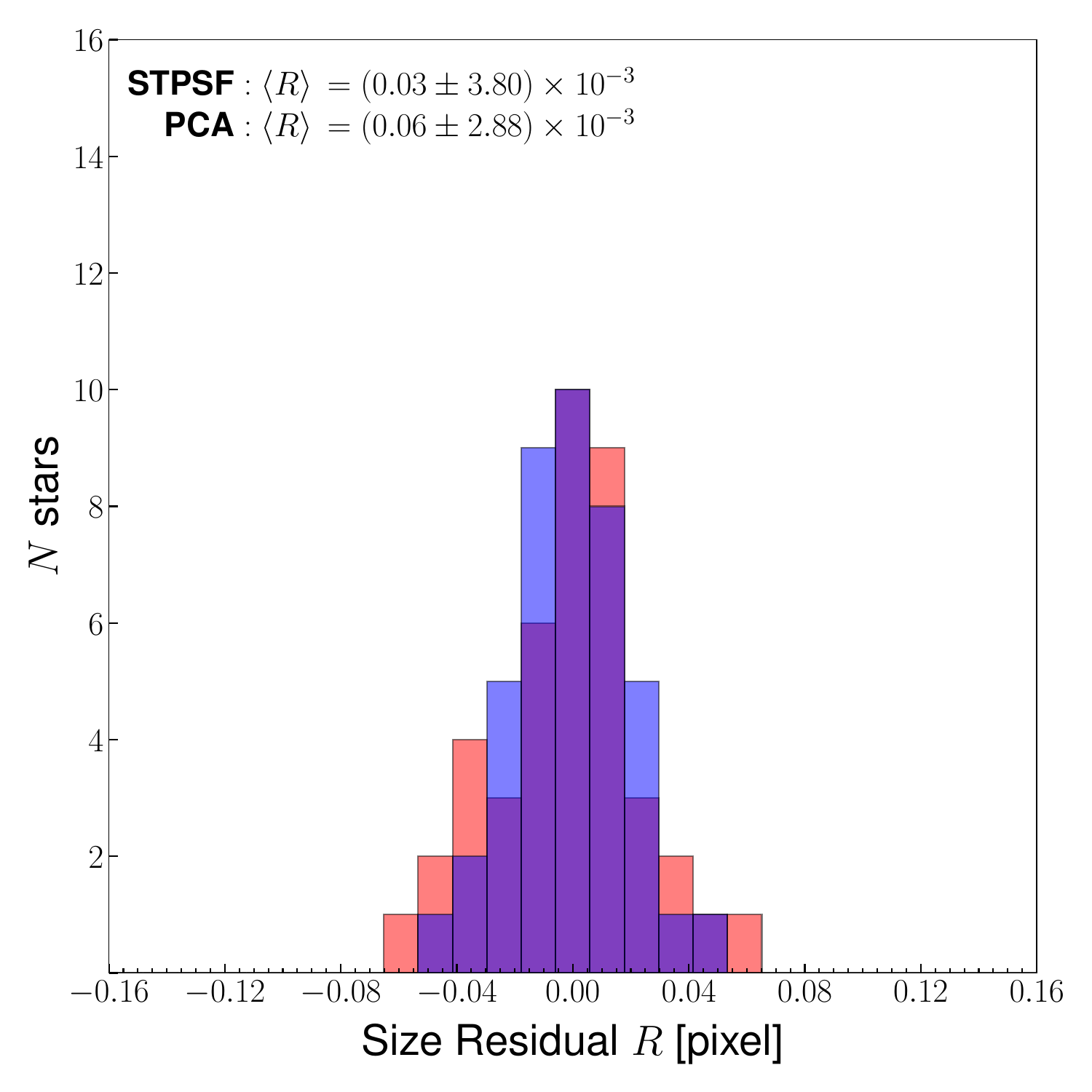}
    
    \centering
    \caption{Residual ellipticity and size measurements for the PCA and \texttt{STPSF} models. \textbf{Top:} Complex ellipticity components of stars measured in the XLSSC~122 field (black circles), and residual \texttt{STPSF} model (red triangles) and PCA model (blue squares) ellipticities computed as the difference between observed and predicted values. The outer circle represents an ellipticity of $|e|=0.02$, while the inner dotted circle corresponds to $|e|=0.01$. \textbf{Bottom:} \texttt{STPSF} (red) and PCA (blue) model residual sizes computed as the difference between observed star sizes and model sizes.}
    \label{fig:psf}
\end{figure}

To create \texttt{STPSF} simulated PSFs to compare \edit{with} the empirical PCA technique, we use the \texttt{stpsf-mosaic} \edit{pipeline} \citep{stpsfmosaic}, which generates PSFs at selected positions while accounting for detector orientations and exposure times.

The residual ellipticity and size measurements for both modeling techniques are shown in Figure \ref{fig:psf}. The top panel presents the residual ellipticities of the \texttt{STPSF} model (red points) and the PCA model (blue points), computed by subtracting the model-predicted ellipticities (measured using quadrupole moments) from the observed star ellipticities (black points) derived using the same method. The \texttt{STPSF} residuals have mean values and standard errors of $\langle e_1 \rangle = (-3.2 \pm 6.0) \times 10^{-4}$ and $\langle e_2 \rangle = (5.6 \pm 7.7) \times 10^{-4}$. For the PCA model, the corresponding values are $\langle e_1 \rangle = (-1.0 \pm 6.7) \times 10^{-4}$ and $\langle e_2 \rangle = (0.4 \pm 7.4) \times 10^{-4}$. The PCA model yields residuals that are slightly more centered around zero with comparable standard errors, indicating a minor but consistent improvement in agreement with the observed star shapes. 

The bottom panel of Figure~\ref{fig:psf} shows the residual PSF sizes as a histogram, using the same color scheme as in the top panel. The \texttt{STPSF} model yields a median and standard error of $\langle R \rangle = (0.03 \pm 3.80) \times 10^{-3}$, while the PCA model gives $\langle R \rangle = (0.06 \pm 2.88) \times 10^{-3}$. This comparison is less informative than the ellipticity residuals, as the \texttt{STPSF} model PSFs are smoothed with an empirically chosen Gaussian kernel to reproduce observed star sizes, which are systematically larger than the simulated PSFs. Together, the residual ellipticity and size measurements indicate that both models effectively reproduce the observed PSF characteristics, with the PCA model performing marginally better, likely due to its data-driven construction. Nevertheless, the close agreement between the two models demonstrates the effectiveness of the \texttt{STPSF} approach and supports its use in star-sparse fields, especially for cluster-scale WL analyses.

\section{\edit{SZ--DM Offset in Cosmological Simulations}}\label{sec:tng}

\begin{figure*}[ht]
    \centering
    \includegraphics[width=\linewidth]{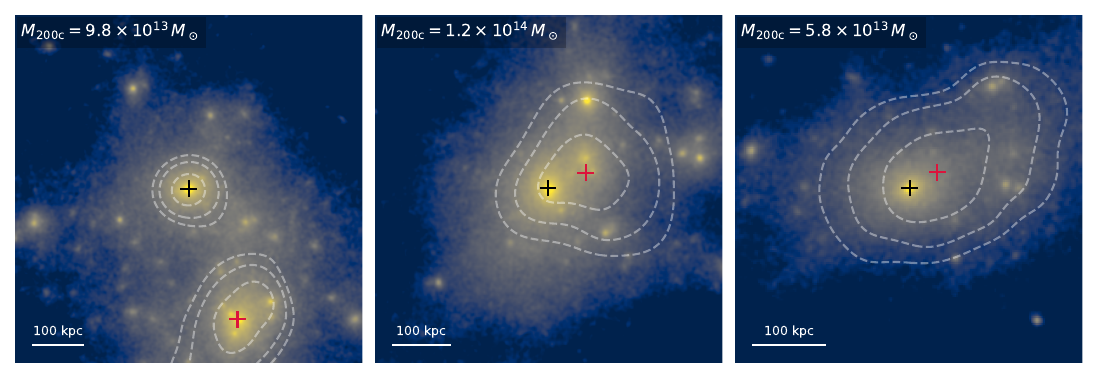}
    \caption{\edit{Projected dark-matter surface mass density ($\Sigma_{\rm DM}$) maps for three representative \textsc{TNG-Cluster} systems illustrating mechanisms producing BCG--SZ offsets. Dashed contours indicate SZ distributions, black crosses mark $\Sigma_{\rm DM}$ peaks (consistent with the BCGs), and red crosses denote SZ peaks. All systems are at $z=1.496$. Left: a pre-merger system with an infalling subhalo. Middle: a post-merger system. Right: a system in which AGN feedback drives the SZ peak displacement.}}
    \label{fig:tng}
\end{figure*}

\edit{Figure~\ref{fig:tng} shows representative examples from the \textsc{TNG-Cluster} simulation highlighting physical mechanisms capable of producing offsets between the BCG and the SZ peak without corresponding BCG--X-ray offsets. The left and middle panels show merger-driven configurations, while the right panel shows a system in which AGN feedback is the primary driver of the SZ peak displacement.}

\edit{In the pre-merger example (left), the SZ peak (red cross) is coincident with an infalling subcluster and offset from the BCG (black cross) by $\sim\!270~\rm kpc$. Offsets of this magnitude, coupled with a clear association between the SZ peak and an infalling subcluster, are commonly observed in pre-merger \textsc{TNG-Cluster} systems. In contrast, XLSSC~122 exhibits an isolated SZ peak $\sim\!100~\rm kpc$ from the BCG.}

\edit{The post-merger configuration shown in the middle panel exhibits an SZ peak that is spatially isolated from dense substructure and displaced by $\sim\!69~\rm kpc$ from the BCG. This morphology is typical of post-merger \textsc{TNG-Cluster} systems and resembles the configuration observed in XLSSC~122.}

\edit{The right panel illustrates a system in which AGN produces a modest SZ peak displacement of $\sim\!43~\rm kpc$ from the BCG. As discussed in Section~\ref{sec:merger}, there is no evidence for AGN activity in XLSSC~122, and AGN-driven configurations in \textsc{TNG-Cluster} typically produce BCG--SZ offsets no larger than $\sim\!50~\rm kpc$.}

\clearpage

\bibliographystyle{aasjournalv7}
\bibliography{main}

\end{document}